# Spin Dephasing and "Hot Spins" **

*Supriyo Datta*

School of Electrical & Computer Engineering,

Purdue University, West Lafayette, IN 47906

email: **datta@purdue.edu**

**Outline**

1. Introduction
2. Hot spin device
3. NEGF equations: A summary
4. NEGF model: Spin valve with impurities
5. Supplementing NEGF: Hot spin effect
6. Summary

**Appendix:** Dephasing in the NEGF formalism

**ABSTRACT**

At this summer school, I have tried to describe a general approach to quantum transport problems based on the Non-equilibrium Green Function (NEGF) method following the approach described in [1]. I will not repeat this treatment here. Instead I will use this article to draw attention to a core issue in quantum transport and how it is modeled within the NEGF method. Specifically, I will discuss what constitutes a dephasing process and how the standard descriptions may need to be extended to include what I will call "hot spin" effects. These involve a number of subtle issues that tend to get hidden in a purely formal description. So I have chosen instead to illustrate the concepts using a concrete device, which is specially engineered to enhance its sensitivity to spin dephasing and hot spin effects.

---

**1. Introduction**

At this summer school, I have tried to describe a general approach to quantum transport problems based on the Non-equilibrium Green Function (NEGF) method following the simple heuristic approach described in [1]. I will not repeat this treatment here. Instead I will use this article to draw attention to a core issue in quantum transport and how it is modeled within the NEGF method. Specifically, I will discuss what constitutes a dephasing process and how the standard descriptions may need to be extended to include what I will call "hot spin" effects in analogy with the more well-known "hot phonon" effects. These involve a number of subtle issues that tend to get hidden in a purely formal description. So I have chosen instead to illustrate the concepts using a concrete device, which is specially engineered to enhance its sensitivity to spin dephasing and hot spin effects. I will call this device a ***"spin-charge transducer"*** for it converts spin excitations into charge currents and vice-versa. It is closely related to the well-known spin-valve device [2], but to my knowledge the spin-charge transducer described here has not been discussed before. I believe devices of this type deserve a serious look from an applied point of view [S.Datta, Proposal for a Spin Capacitor, Appl. Phys. Lett. **87** (2005) 013115], but that is a different story - here I will use it purely as a pedagogical tool.

*NEGF method:* Before I get into dephasing, let me say a few words about the general approach to quantum transport based on the NEGF method. Fig.1a shows a schematic illustration of a nanoscale transistor structure which could be used as a nanosensor as well if its current-voltage (I-V) characteristics were changed significantly by a foreign object M. The NEGF method, illustrated schematically in Fig.1b and summarized in Section 3, provides a general approach for modeling this entire class of devices and has been applied by different authors to many different kinds of devices like semiconductor nanowires, carbon nanotubes and molecular wires, having many different kinds of contacts, both normal and superconducting. Many issues still need to be clarified such as the best choice of basis functions in different problems, but I think it is safe to say that this approach provides a suitable framework for a wide variety of problems involving nanoscale quantum transport.

In modeling the electrical conduction through such nanoscale structures it is common to to assume that transport through the active region (marked "channel" in Fig.1a) is completely "coherent". "Incoherent"processes are assumed to occur only in the contacts (labeled "source" and "drain" in Fig.1a) where they serve to maintain local equilibrium. Current flow in real devices, especially at large bias voltages inevitably involve significant amounts of "incoherent" processes within the channel and it is natural to ask what distinguishes them from coherent processes and how they can be included in the model. This is the question I would like to discuss in this article.

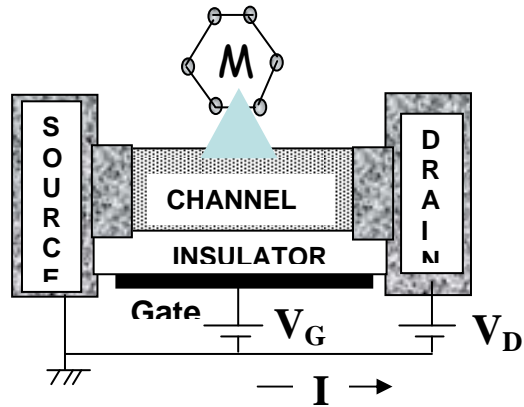

Fig.1.1a. Schematic illustration of a nanotransistor / nanosensor. "M" denotes a foreign object that could be sensed through its effect on the current.

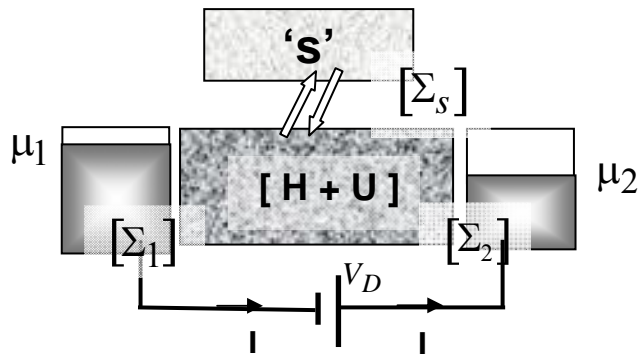

Fig.1.1b. Generic model for quantum transport (adapted from [1]). The meanings of different symbols are discussed in Section 3.

***What constitutes a dephasing process?*** So what constitutes "dephasing"? Or in other words, what distinguishes a coherent from an incoherent process? The standard definition of coherent transport is a process in which the passage of an electron from the source to the drain does not change the state of any other entity. If the electron were simply deflected by a rigid impurity in the channel, it would constitute a coherent process. But if it were to deliver some energy to the atoms and set them jiggling (in other words, excite a phonon) it would constitute an incoherent process since the state of the atoms is changed. This suggests that incoherent processes should also be inelastic, that is, involve an exchange of energy. But that is not necessarily true and spin dephasing is a classic example of that. Imagine a magnetic impurity with two spin states of equal energy (degenerate), one up and one down such that an electron can interact with it and flip it from one state to the other. No energy is exchanged making the process an elastic one. Nevertheless, it is an incoherent process since the state of the impurity has changed.

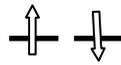

But what makes such processes incoherent? Could we not consider the electron and the impurity spin as one composite system and claim the process to be coherent, since the state of nothing else has changed? What really makes such processes incoherent is that external forces conspire to return the impurity spins to an unpolarized (50% up, 50% down) distribution, or the phonons to a Bose-Einstein distribution characteristic of thermal equilibrium. The process by which equilibrium is restored is seldom discussed explicitly: It is assumed to be fast enough to maintain the impurity spins (or the jiggling atoms) in a state of thermal equilibrium. Its effect enters the model somewhat surreptitiously through an innocent-looking "boundary condition". And yet it is this process of "information erasure" that is the essence of dephasing.

To illustrate this point, I will use a (gedanken) "spin-charge transducer" that is specially engineered to sensitize the terminal current to spin-flip interactions with impurity spins and to the impurity spin polarization. As long as external forces are able to maintain the impurity spins in an unpolarized state, they are effective in flipping electronic spins resulting in a large terminal current. But these spin-flip interactions can soon overwhelm the external forces and polarize the impurities so that they are no longer effective in flipping electronic spins, causing a reduction in the terminal current. We will refer to this as the "hot spin" effect in analogy with the hot phonon effect that arises when the phonon emission rate overwhelms the phonon extraction mechanisms. I believe that the hot spin effect could arise even at voltages as low as tens of meV because the impurity spin restoration mechanisms are typically quite weak and easily overwhelmed. This would be even more true, if nuclear spins [3] were the dephasing agents: The basic physics is much the same, it is just that the spin restoration time is much longer.

The operation of this "spin-charge transducer" can be understood in terms of a simple equivalent circuit which I will introduce heuristically in **Section 2** and justify in **Section 4** using the NEGF model summarized in **Section 3**. Finally in **Section 5**, I will show how the hot spin effect requires us to supplement the NEGF method with a dynamical equation for the impurity spin. This effect may have interesting practical applications such as a spin-based "flash memory", but we will not get into this aspect at all. My objective is to illustrate the assumptions that underlie our models for dephasing processes and how these models may need to be modified when these assumptions are violated in future nanoscale devices.

## 2. Proposed "spin-charge" transducer

In this section I will describe the basic device structure that I will use for illustrating dephasing effects in spin devices. The idea is based on the spin valve [2] where the current is reduced significantly when an external magnetic field switches the contacts from the parallel (P) configuration to the anti-parallel (A) configuration (Fig.2.1a). This switching is typically accomplished by making one contact "softer", so that it switches its magnetization at a lower magnetic field than the other. The resulting magnetoresistance can be useful in "reading" information stored in magnetic disks.

The reduction in the current through a spin valve from the "P" to the "A" configuration can be understood intuitively from a simple circuit model (Fig.2.1b), which we will justify quantitatively from the NEGF formalism in Section 4. In this circuit model, we associate a contact resistance of $g_\alpha^{-1}$ with the majority spin and a different one $g_\beta^{-1}$ with the minority spin. It is then straightforward to write the measured terminal conductances in the 'P' and 'A' configurations denoted by $G_P$ and $G_A$ in terms of the majority and minority contact conductances $g_\alpha$ and $g_\beta$:

$$G_P = (g_\alpha + g_\beta)/2 \quad , \qquad G_A = 2 g_\alpha g_\beta /(g_\alpha + g_\beta) \tag{2.1}$$

Defining the contact polarization $P_C$ as

$$P_C \equiv (g_\alpha - g_\beta)/(g_\alpha + g_\beta) \tag{2.2}$$

we can write the ratio of the conductances in the "A" and "P" configurations as

$$G_A / G_P = 1 - P_C^2 \tag{2.3}$$

***Spin valve with magnetic impurities:*** Consider now a spin valve device with magnetic impurities in the channel that cause spin-flip scattering. This is an example of a dephasing process and in Section 4 we will see how they are described within the NEGF formalism. However, we can get a simple circuit model for spin-flip processes by including a spin-flip conductance $g_\gamma$ bridging the up-spin and down-spin channels. It is straightforward to show that the terminal conductances in the 'P' and 'A' configurations denoted by $G_P$ and $G_A$ are now changed from Eq.(2.1) to

$$G_P = (g_\alpha + g_\beta)/2 \tag{2.4a}$$

$$G_A = \frac{(g_\alpha^2 + g_\beta^2)g_\gamma + 2g_\alpha g_\beta (g_\alpha + g_\beta + g_\gamma)}{(g_\alpha + g_\beta)(g_\alpha + g_\beta + 2g_\gamma)} \tag{2.4b}$$

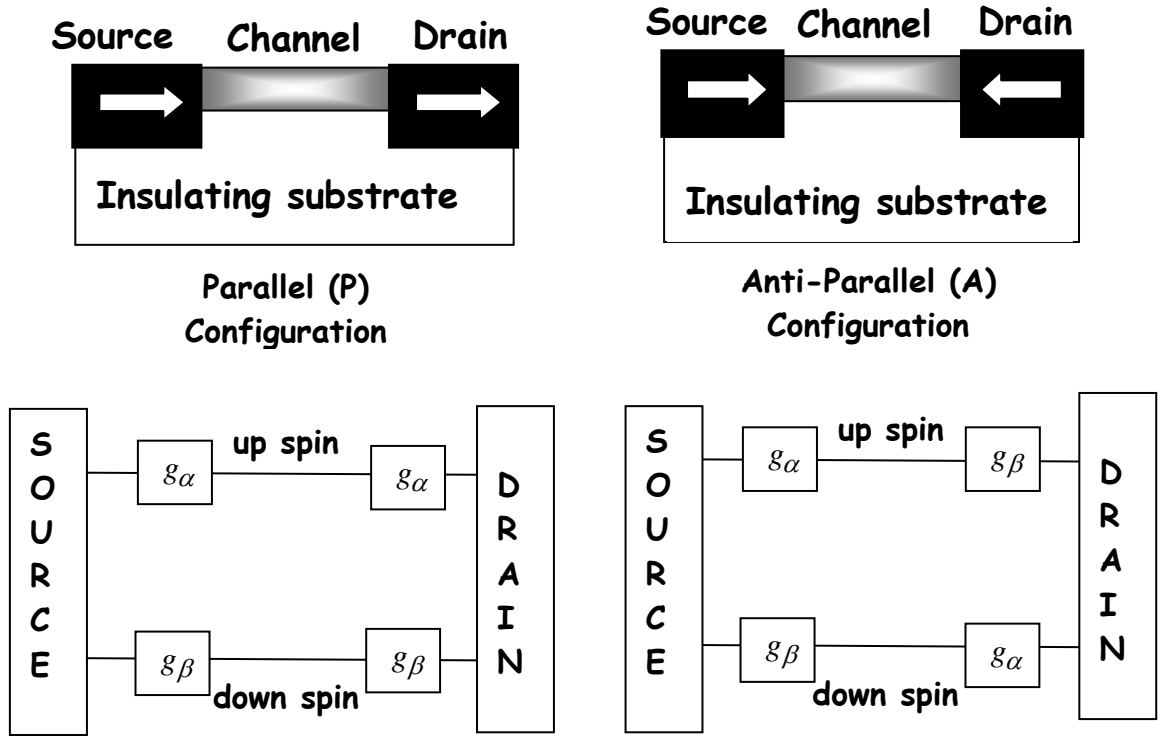

Fig.2.1. (a) Generic spin-valve device showing parallel (P) and anti-parallel (A) configurations. (b) Simple intuitive equivalent circuit for each configuration.

Note that the parallel conductance $G_P$ is independent of spin-flip scattering because the spin-flip conductance $g_\gamma$ bridges two points that are exactly at the same potential halfway between the two contacts (see Fig.2b). This of course may not be precisely true in real devices where the two contacts may not be identical and may require four separate contact conductances for detailed modeling, rather than just $g_\alpha$ and $g_\beta$ as we have assumed (an assumption that is easily relaxed) for simplicity.

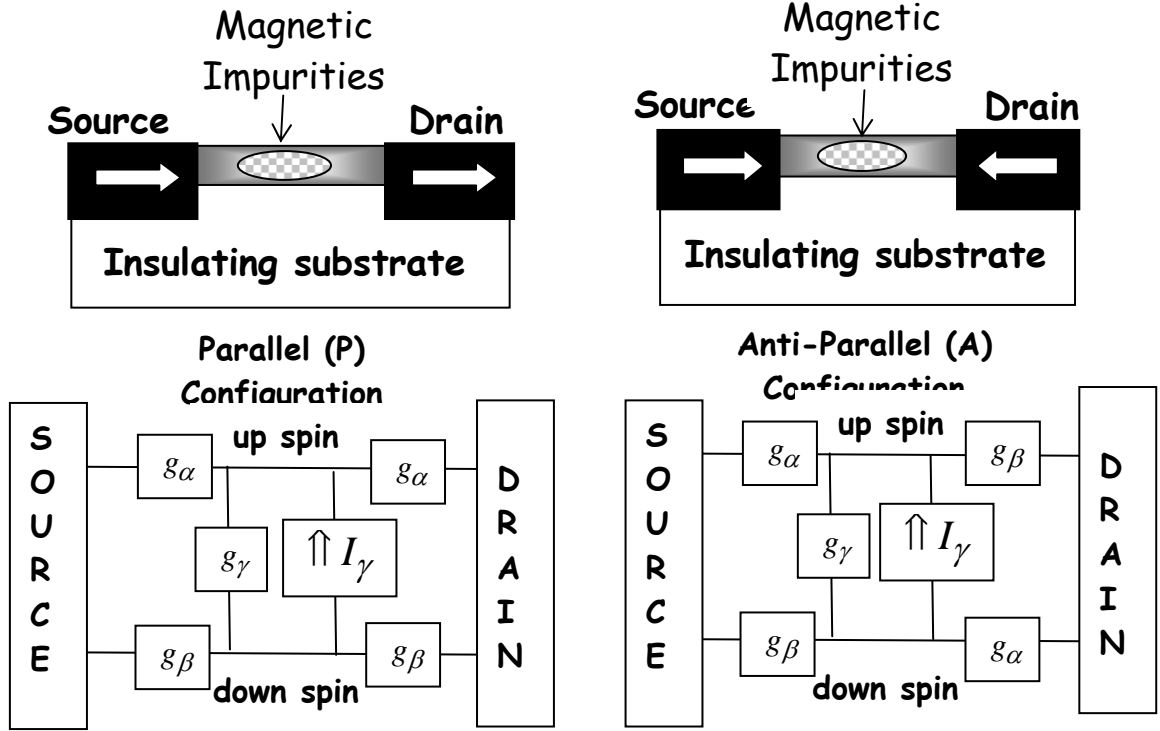

Fig.2.2. Proposed "spin-charge transducer": Same as Fig.2.1 but with magnetic impurities present in the channel which in the simplest case can be modeled as a "spin-flip conductance $g_\gamma$" bridging the upspin and downspin channels. More generally one needs a current generator $I_\gamma \sim (F_u - F_d)$, where $F_u$ and $F_d$ are the fraction of impurities with their spins up and down respectively.

Using Eq.(2.4) we write the conductance ratio as

$$\frac{G_A}{G_P} = 1 - \frac{g_\alpha + g_\beta}{g_\alpha + g_\beta + 2g_\gamma} P_C^2 \qquad (2.5)$$

which reduces to Eq.(2.3) if the spin-flip conductance $g_\gamma$ is much less than the contact conductance ($g_\alpha + g_\beta$). It is evident that the effect of spin-flip impurities is to make the conductance ratio $G_A/G_P$ closer to one as we might expect. Any mechanism that randomizes the spin orientation within the channel serves to wipe out the distinction between the "A" and "P" configurations and brings the conductance ratio closer to one. However, one of the key concepts I wish to stress is that spin-flip impurities do not necessarily randomize the spin as is commonly assumed. Let me explain.

*Hot spin effect:* Spin-flip impurities randomize the spin only if there are external forces constantly maintaining them in an equilibrium state with the fraction of impurities, $F_u$ that point up equal to the fraction of impurities, $F_d$ that point down. But if the impurities are weakly coupled to the surroundings (other than the conduction electrons) then the impurities will reach a steady state polarization with $F_u \neq F_d$ such that there is no further spin-flip scattering. As a result, the conductance ratio $G_A/G_P$ might start out at the value given by Eq.(2.5) but will eventually reach the value given by Eq.(2.3) as the impurities get polarized by the flow of current. In other words the steady-state conductance ratio $G_A/G_P$ will remain **unchanged by the presence of the spin-flip impurities**!

We could include this effect in the circuit model in Fig.2.1 through a current generator $I_\gamma \sim (F_u - F_d)$ in parallel with the conductance $g_\gamma$ (Fig.2.2). If we start with randomly oriented impurities such that $F_u = F_d$, then the current generator is zero and we obtain Eq.(2.5). But if the impurities have no independent relaxation mechanism then a steady state polarization ($F_u - F_d$) will develop such that the current generator exactly cancels the current through $g_\gamma$, so that we get back the same result Eq.(2.3) that we obtain without $g_\gamma$ (see Fig.2.1).

Note that in the ideal "P" configuration (Fig.2.2), this dynamic polarization of impurities is bias-independent and has no effect on the terminal current (see Eq.(2.4a)), although there could be some residual effect in real structures with non-identical contacts. But by setting up the device *in the 'A' configuration* we make the impurity polarization bias-dependent and clearly observable through its effect on the terminal current. For example if we short circuit the external terminals, any existing $I_\gamma$ will not lead to any short circuit current in the external circuit in the "P" configuration, but will lead to a current of

$$I_{sc} = I_\gamma (g_\alpha - g_\beta)/(2g_\gamma + g_\alpha + g_\beta) \qquad (2.6)$$

in the "A" configuration.

Before we move on, let me summarize by noting that the "spin-charge transducer" is simply a ***spin valve with anti-parallel contacts*** containing ***spin-flip impurities***. What makes this device unique is the sensitivity of the terminal current to the degree of spin-flip scattering. As long as external forces can maintain the impurities in an unpolarized state, they serve as spin-flip scatterers and a large current flows in the external circuit. But if the impurities cannot relax easily, then they can get polarized and cease to act as spin-flip scatterers, with a reduction in the terminal current.

### 3. NEGF Equations : A summary

As I mentioned in the introduction, a fairly versatile model for quantum transport in nanoscale devices is now available based on the Non-equilibrium Green Function (NEGF) method [1]. In this section let me summarize the basic NEGF equations.

***Inputs to the NEGF model:*** The inputs (see Fig.1.1b) can be classified in four categories:

***1. Channel:*** The channel or active region is described by a ***Hamiltonian*** matrix, [H], which also includes the Laplace potential $[U_L]$ due to the applied voltages on the electrodes.

***2. Contacts:*** The source and drain contacts are described by the ***self-energy matrices***, $[\Sigma_{1,2}(E)]$. The corresponding ***broadening*** matrices are given by their anti-Hermitian components:

$$\Gamma_{1,2}(E) = i[\Sigma_{1,2}(E) - \Sigma_{1,2}^+(E)] \qquad (3.1)$$

while the ***inscattering*** matrices are equal to the broadening matrix times the corresponding Fermi functions f(E):

$$[\Sigma_1^{in}(E)] = f_1(E)[\Gamma_1(E)] \qquad (3.2a)$$

$$[\Sigma_2^{in}(E)] = f_2(E)[\Gamma_2(E)] \qquad (3.2b)$$

and the *outscattering* matrices are equal to the broadening matrix times (1-f):

$$[\Sigma_1^{out}(E)] = (1 - f_1(E))[\Gamma_1(E)] \qquad (3.3a)$$

$$[\Sigma_2^{out}(E)] = (1 - f_2(E))[\Gamma_2(E)] \qquad (3.3b)$$

Note that the Fermi functions in the two contacts are given by

$$f_1(E) = \frac{1}{1 + \exp((E - \mu_1)/k_B T)} \qquad (3.4a)$$

and

$$f_2(E) = \frac{1}{1 + \exp((E - \mu_2)/k_B T)} \qquad (3.4b)$$

where $\mu_1 = E_f + (eV_D/2)$ and $\mu_2 = E_f - (eV_D/2)$, $E_f$ being the equilibrium Fermi energy and $V_D$ being the applied drain voltage (relative to the source).

**3. Charging:** The charging effects of a change in the number of electrons in the channel are included through the *potential* matrix [U] describing the effective potential that one electron feels.

**4. Scattering:** Incoherent scattering processes due to the interaction of the channel with the surroundings are described by the *inscattering* $[\Sigma_s^{in}(E)]$ ($\equiv [-i\Sigma^<(E)]$) *and outscattering* $[\Sigma_s^{out}(E)]$ ($\equiv [+i\Sigma^>(E)]$) *matrices.* The corresponding broadening matrix $\Gamma_s(E)$ is the sum of the two (using the subscript 's' to distinguish the scattering or interaction induced broadening from the contact induced broadenings in Eq.(3.1))

$$\Gamma_s(E) = [\Sigma_s^{in}(E) + \Sigma_s^{out}(E)] \qquad (3.5a)$$

and is equal to half the imaginary part of the self-energy matrix $\Sigma_s(E)$:

$$\Sigma_s(E) = P\left(\int \frac{dE'\,\Gamma(E')/2\pi}{E'-E}\right) - i\,\frac{\Gamma(E)}{2} \qquad (3.5b)$$

Note that the real part of the self-energy matrix is the Hilbert transform of the imaginary part as required of the Fourier transform of any causal function.

***NEGF Equations:*** Given these inputs, the NEGF equations tell us how to calculate the Green function [G(E)],

$$G = [EI - H - U - \Sigma]^{-1} \quad \textbf{with} \quad \Sigma = \Sigma_1 + \Sigma_2 + \Sigma_s \qquad (3.6)$$

the ***spectral function*** [A(E)] (analogous to density of states)

$$A = i[G - G^+] \qquad (3.7)$$

and the ***correlation function*** $[G^n(E)](\equiv [-iG^<(E)])$ ( analogous to the electron density)

$$G^n = G\,\Sigma^{in}\,G^+ \quad \textbf{with} \quad \Sigma^{in} = \Sigma_1^{in} + \Sigma_2^{in} + \Sigma_s^{in} \qquad (3.8)$$

respectively. The ***hole correlation function*** $[G^p(E)](\equiv [+iG^>(E)])$ (analogous to the hole density) can be calculated from a similar relation

$$G^p = G\,\Sigma^{out}\,G^+ \quad \textbf{with} \quad \Sigma^{out} = \Sigma_1^{out} + \Sigma_2^{out} + \Sigma_s^{out} \qquad (3.9)$$

or alternatively from a knowledge of $[G^n(E)]$ and $[A(E)]$ by invoking the relation

$$A = G^n + G^p \tag{3.10}$$

which can be shown to be consistent with Eq.(3.6).

From a knowledge of the spectral and correlation functions, any one-electron observable quantity can be obtained. For example, the current (per spin) at any terminal 'i' can be calculated from

$$I_i = \int_{-\infty}^{+\infty} dE \; \tilde{I}_i(E)$$

$$\text{with} \quad \tilde{I}_i = (q/h)(Trace\,[\Sigma_i^{in} A] - Trace\,[\Gamma_i G^n]) \tag{3.11}$$

**4. NEGF model: Spin valve with impurities**

To apply the NEGF approach to the device proposed in Section 2, we first need the input matrices listed at the start of section 3. There is an important distinction between the inputs in categories 1,2 (channel and contacts) and those in categories 3,4 (charging and scattering). The inputs in categories 1 and 2 are specified and fixed from the outset in any calculation. By contrast, the inputs [U], [$\Sigma^{in}$] and [$\Sigma^{out}$] under categories 3 and 4 depend on the correlation function and spectral function obtained from Eqs.(3.4) through (3.6) and as such have to be calculated *self-consistently* using an *iterative* procedure. Let us go through each of these categories:

*Channel:* All the matrices involved are generally of size (NxN), N being the number of basis functions needed to represent the channel. To keep things simple, I will assume that the channel is described by just two basis functions (N=2), one for upspin and one for downspin and write the Hamiltonian as

$$H = \begin{bmatrix} \varepsilon & 0 \\ 0 & \varepsilon \end{bmatrix} \tag{4.1}$$

assuming zero off-diagonal elements. Note that a channel with spin-orbit coupling would be described by a Hamiltonian with off-diagonal elements like

$$H = \begin{bmatrix} \varepsilon & \Delta \\ \Delta^* & \varepsilon \end{bmatrix}$$

The general NEGF formalism can easily handle such off-diagonal elements and this is needed to handle many of the proposed device concepts. But here I will restrict myself to the simplest case where all the relevant matrices are diagonal. The resulting NEGF equations under these conditions can be viewed as simple rate equations that follow from much more elementary arguments [1]. As such the full power of the NEGF equations is not really needed, but my purpose is to illustrate the underlying physics while minimizing distracting details.

*Contacts:* It is convenient to visualize each of the contacts (source and drain) as two separate contacts, one for upspin electrons and one for downspin electrons, each coupled to the channel by its own (2x2) broadening matrix assuming that the two magnetic contacts are in the "A" configuration:

$$\text{Source:} \quad \Gamma_{1,u} = \begin{bmatrix} \alpha & 0 \\ 0 & 0 \end{bmatrix}, \quad \Gamma_{1,d} = \begin{bmatrix} 0 & 0 \\ 0 & \beta \end{bmatrix}$$

$$\text{Drain:} \quad \Gamma_{2,u} = \begin{bmatrix} \beta & 0 \\ 0 & 0 \end{bmatrix}, \quad \Gamma_{2,d} = \begin{bmatrix} 0 & 0 \\ 0 & \alpha \end{bmatrix} \quad (4.2)$$

. The inscattering functions from the contacts are given by (see Eqs.(3.2a,b))

$$\Sigma_{1,u}^{in} = \begin{bmatrix} f_1\alpha & 0 \\ 0 & 0 \end{bmatrix}, \quad \Sigma_{1,d}^{in} = \begin{bmatrix} 0 & 0 \\ 0 & f_1\beta \end{bmatrix}$$

$$\Sigma_{2,u}^{in} = \begin{bmatrix} f_2\beta & 0 \\ 0 & 0 \end{bmatrix}, \quad \Sigma_{2,d}^{in} = \begin{bmatrix} 0 & 0 \\ 0 & f_2\alpha \end{bmatrix} \quad (4.3)$$

***Spin-flip interactions:*** We now have the inputs listed under categories 1 and 2 at the start of Section 3 which are specified and fixed from the outset of any calculations. By contrast, the inputs [U], [$\Sigma^{in}$] and [$\Sigma^{out}$] under categories 3 and 4 depend on the correlation function and spectral function obtained from Eqs.(3.4) through (3.6) and as such have to be calculated ***self-consistently*** using an ***iterative*** procedure. I will neglect [U] which is justified if the spectral function is approximately constant over the energy range of interest. This allows us to focus on item 4, specifically, the dephasing due to the spin-flip interaction with magnetic impurites.

Scattering processes in general are described by one fourth-order tensor, $D^n$ relating the inscattering matrix to the correlation function

$$\Sigma^{in}_{ij} \;=\; \sum_{k,l} D^n_{ijkl} \, G^n_{kl} \tag{4.4a}$$

and another $D^p$ relating the outscattering matrix to the hole correlation function

$$\Sigma^{out}_{ij} \;=\; \sum_{k,l} D^p_{ijkl} \, G^p_{kl} \tag{4.4b}$$

Each of these quantities $D^n$ and $D^p$ in general have 16 components relating each of the four components of the correlation functions with each of the four components of the scattering function. But if we restrict ourselves to problems where the off-diagonal elements of the correlation functions are all zero, then we need only four of these components to connect the diagonal elements. If we assume that the interaction between the electron spin $\vec{\sigma}$ and the impurity interaction $\vec{S}$ is written as (J: exchange coupling between a channel electron and an impurity)

$$H_{\text{int}} \;=\; J\vec{\sigma}.\vec{S} \tag{4.5}$$

then
$$\begin{Bmatrix} \Sigma^{in}_s(1,1) \\ \Sigma^{in}_s(2,2) \end{Bmatrix} \;=\; J^2 N_I \begin{bmatrix} 1 & 4F_u \\ 4F_d & 1 \end{bmatrix} \begin{Bmatrix} G^n(1,1) \\ G^n(2,2) \end{Bmatrix} \tag{4.6a}$$

$$\begin{Bmatrix} \Sigma_s^{out}(1,1) \\ \Sigma_s^{out}(2,2) \end{Bmatrix} = J^2 N_I \begin{bmatrix} 1 & 4F_d \\ 4F_u & 1 \end{bmatrix} \begin{Bmatrix} G^p(1,1) \\ G^p(2,2) \end{Bmatrix} \quad (4.6b)$$

where $N_I$ is the number of magnetic impurities, and $F_u$ and $F_d$ represent the fraction of these impurities that have their spins up and down respectively ($F_u + F_d = 1$). A formal derivation of Eqs.(4.6) including the (12) and (21) components as well is provided in Appendix A. But the expressions for the (11) and (22) components given above can be understood heuristically. For example, the inscattering into the upspin component $\Sigma_s^{in}(11)$ is proportional to the density of down spin electrons $G^n(22)$ times the number of upspin impurities, $N_I F_u$ etc.

We could denote the (1,1) component with a subscript 'u' for upspin and the (2,2) component with 'd' for downspin to write:

$$\Sigma_u^{in} = J^2 N_I (G_u^n + 4F_u G_d^n) \quad (4.7a)$$

$$\Sigma_d^{in} = J^2 N_I (G_d^n + 4F_d G_u^n) \quad (4.7b)$$

$$\Sigma_u^{out} = J^2 N_I (G_u^p + 4F_d G_d^p) \quad (4.8a)$$

$$\Sigma_d^{out} = J^2 N_I (G_d^p + 4F_u G_u^p) \quad (4.8b)$$

so that from Eq.(3.5a) making use of Eq.(3.8),

$$\Gamma_u = J^2 N_I (A_u + 4F_u G_d^n + 4F_d G_d^p) \quad (4.9a)$$

$$\Gamma_d = J^2 N_I (G_d^n + 4F_d G_u^n + 4F_u G_u^p) \quad (4.9b)$$

Note that our treatment of spin-flip scattering is based on the assumption that the impurity density matrix is diagonal: $\begin{bmatrix} F_u & 0 \\ 0 & F_d \end{bmatrix}$. With off-diagonal elements present one would need the other terms of the tensors $D^n$ and $D^p$ as well (see Appendix A for a general discussion). The assumption of zero off-diagonal elements in the (1) impurity density matrix, (2) the channel Hamiltonian and (3) the contact self-energies ensure that the resulting equations involve only the diagonal elements and have simple common sense interpretations. This allows us to understand the formalism in simple common sense terms, but it can be applied to more general problems involving off-diagonal terms.

*Current:* From Eqs.(3.7) we can write the terminal currents at any energy as

$$\tilde{I}_{1,u} = (e/h)\alpha A_u (f_1 - f_u), \quad \tilde{I}_{1,d} = (e/h)\beta A_d (f_1 - f_d) \quad (4.10a)$$

$$\tilde{I}_{2,u} = (e/h)\beta A_u (f_2 - f_u), \quad \tilde{I}_{2,d} = (e/h)\alpha A_d (f_2 - f_d) \quad (4.10b)$$

where $A_u(E)$ and $A_d(E)$ are the (1,1) and (2,2) elements of the spectral function [A(E)] which represent the density of states (times $2\pi$) associated with the up and down spin levels respectively. The functions $f_u(E)$ and $f_d(E)$ are defined as the ratio of the diagonal elements of the correlation function (which represent the electron density times $2\pi$) to the diagonal elements of the spectral function:

$$f_u(E) = G_u^n(E)/A_u(E) \quad \text{and} \quad f_d(E) = G_d^n(E)/A_d(E) \quad (4.11)$$

*Green function, spectral function and the correlation function:* Substituting the (2x2) matrices introduced in this section into the general NEGF equations (3.6) through (3.8), we can write

$$G_u = \frac{1}{E - \varepsilon + i(\alpha + \beta + \Gamma_u)/2} \qquad G_d = \frac{1}{E - \varepsilon + i(\alpha + \beta + \Gamma_d)/2} \quad (4.12)$$

$$A_u = \frac{\alpha + \beta + \Gamma_u}{(E - \varepsilon)^2 + ((\alpha + \beta + \Gamma_u)/2)^2} \qquad A_d = \frac{\alpha + \beta + \Gamma_d}{(E - \varepsilon)^2 + ((\alpha + \beta + \Gamma_d)/2)^2} \quad (4.13)$$

$$G_u^n = |G_u|^2 (\alpha f_1 + \beta f_2 + \Sigma_u^{in}) \qquad G_d^n = |G_d|^2 (\beta f_1 + \alpha f_2 + \Sigma_d^{in}) \quad (4.14)$$

so that we can write
$$f_u \equiv \frac{G_u^n}{A_u} = \frac{\alpha f_1 + \beta f_2 + \Sigma_u^{in}}{\alpha + \beta + \Gamma_u} \quad (4.15a)$$

and
$$f_d \equiv \frac{G_d^n}{A_d} = \frac{\beta f_1 + \alpha f_2 + \Sigma_d^{in}}{\alpha + \beta + \Gamma_d} \quad (4.15b)$$

***Spin-flip current:*** We can rewrite Eqs.(4.15a,b) in the form

$$\alpha(f_1 - f_u) + \beta(f_2 - f_u) + (\Sigma_u^{in} - \Gamma_u f_u) = 0$$
$$\beta(f_1 - f_d) + \alpha(f_2 - f_d) + (\Sigma_d^{in} - \Gamma_d f_d) = 0$$

and make use of Eq.(4.10a,b) to write

$$\tilde{I}_{1,u} + \tilde{I}_{2,u} + \tilde{I}_{s,u} = 0 \quad (4.16a)$$

and
$$\tilde{I}_{1,d} + \tilde{I}_{2,d} + \tilde{I}_{s,d} = 0 \quad (4.16b)$$

by ***defining***
$$\tilde{I}_{s,u} \equiv (e/h)(\Sigma_u^{in} A_u - \Gamma_u G_u^n)$$

and
$$\tilde{I}_{s,d} \equiv (e/h)(\Sigma_d^{in} A_d - \Gamma_d G_d^n)$$

Substituting for the inscattering and broadening from Eqs.(4.7a,b) and (4.9a,b) respectively and making use of Eq.(4.11) we obtain

$$\tilde{I}_{s,u} = -\tilde{I}_{s,d} = 4(e/h)J^2 N_I A_u A_d (F_u f_d (1 - f_u) - F_d f_u (1 - f_d)) \quad (4.17)$$

Eqs.(4.16a,b) look like Kirchhoff's laws applied to the 'u' and 'd' nodes suggesting a simple physical picture (see Fig.4.1) for the equations that we have derived from the general NEGF model. Indeed, we could have written down the current equations Eqs.(4.10a,b) intuitively [1]. Even Eq.(4.17) can be written down by applying the golden rule to the electron-impurity interaction (Eq.(4.5)). The main thing that the NEGF adds in this context is the notion that the spectral functions $A_u$ and $A_d$ in general have to be calculated self-consistently from Eq.(4.13) and could be modified significantly if the scattering induced broadening $\Gamma_u$ and $\Gamma_d$ are comparable to the fixed contact-induced broadening ($\alpha + \beta$). Of course the real value of the NEGF formalism is that it allows one to handle more complicated spin transport problems involving off-diagonal terms as well.

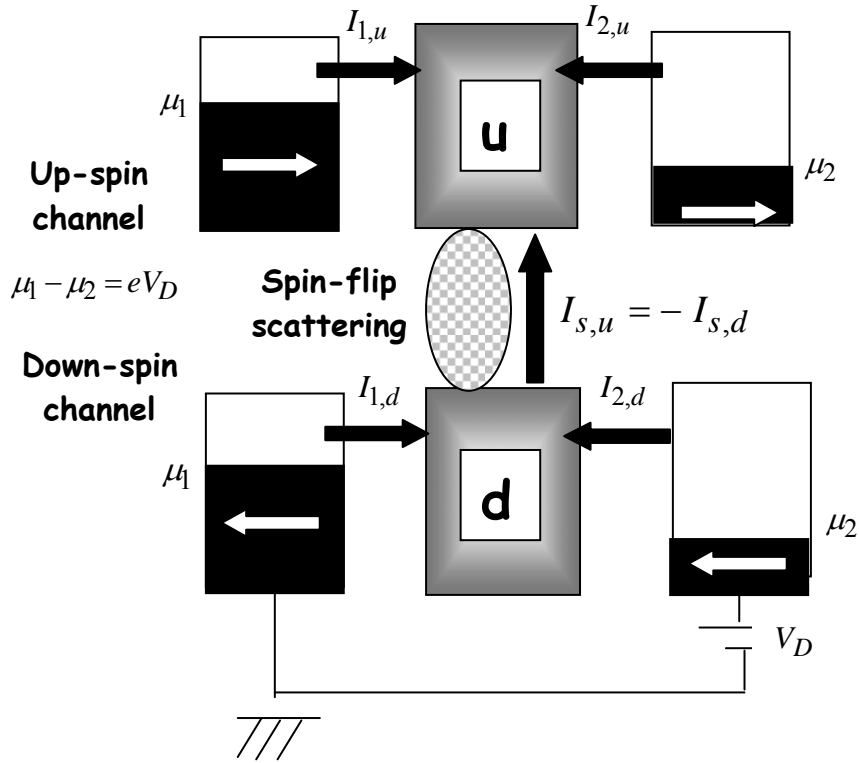

Fig.4.1. Simple physical picture for the NEGF model for a spin-valve with magnetic impurities.

To obtain numerical results, we need to calculate $f_u$ and $f_d$ by solving Eqs.(4.17a,b) and then substitute back into the current equations Eqs.(4.11a,b). Note that while the terminal currents are linear in the occupation factors 'f', the spin-flip current in Eq.(4.18) has quadratic terms. It is convenient to separate the spin-flip current into a linear and a quadratic term:

$$\tilde{I}_{s,u} = -\tilde{I}_{s,d} = \tilde{I}_s + \tilde{I}_\gamma \qquad (4.18a)$$

given by (note that $F_d + F_u = 1$)

$$\tilde{I}_s = 2(e/h)J^2 N_I A_u A_d [f_d - f_u] \qquad (4.18b)$$

$$\tilde{I}_\gamma = 2(e/h)J^2 N_I A_u A_d [f_d(1-f_u) + f_u(1-f_d)](F_u - F_d) \qquad (4.18c)$$

Unlike $\tilde{I}_\gamma$, $\tilde{I}_s$ is linear and looks just like the terminal currents.

***NEGF vs. Equivalent Circuit:*** How do the results from our NEGF model match those from the heuristic equivalent circuit discussed in Section 2? For the moment, let us assume that the impurity spins are in an unpolarized state with $F_d = F_u = 0.5$, so that $\tilde{I}_\gamma = 0$ and the spin-flip current is described entirely by the linear term $\tilde{I}_s$.

In this case it is easy to obtain the equivalent circuit introduced heuristically in Section 2 (see Fig.2.2, without the current generator $I_\gamma$). For this we need the total current which is obtained by integrating the currents given by Eqs.(4.8a,b) over energy. For example,

$$I_{1,u} = (e\alpha/h) \int dE\, A_u (f_1 - f_u)$$

Assuming that the spectral density, $A_u$ is constant over the energy range of interest and that $f_u$ is also a Fermi function with an electrochemical potential $\mu_u$, we can write

$$I_{1,u} = (e\alpha/h) A_u (\mu_1 - \mu_u) \rightarrow (e^2\alpha/h) A_u (V_1 - V_u)$$

suggesting that the conductances appearing in the circuit model of Figs.2.1, 2.2 are given by

$$g_\alpha = (e^2/h) A\alpha \quad \text{and} \quad g_\beta = (e^2/h) A\beta \qquad (4.19a)$$

assuming $A_u = A_d = A$. Similarly the spin-flip conductance $g_\gamma$ is obtained from Eq.(4.18b) and can be written in the same form:

$$g_\gamma = (e^2/h) A \gamma \quad \text{where} \quad \gamma \equiv 2J^2 A N_I \qquad (4.19b)$$

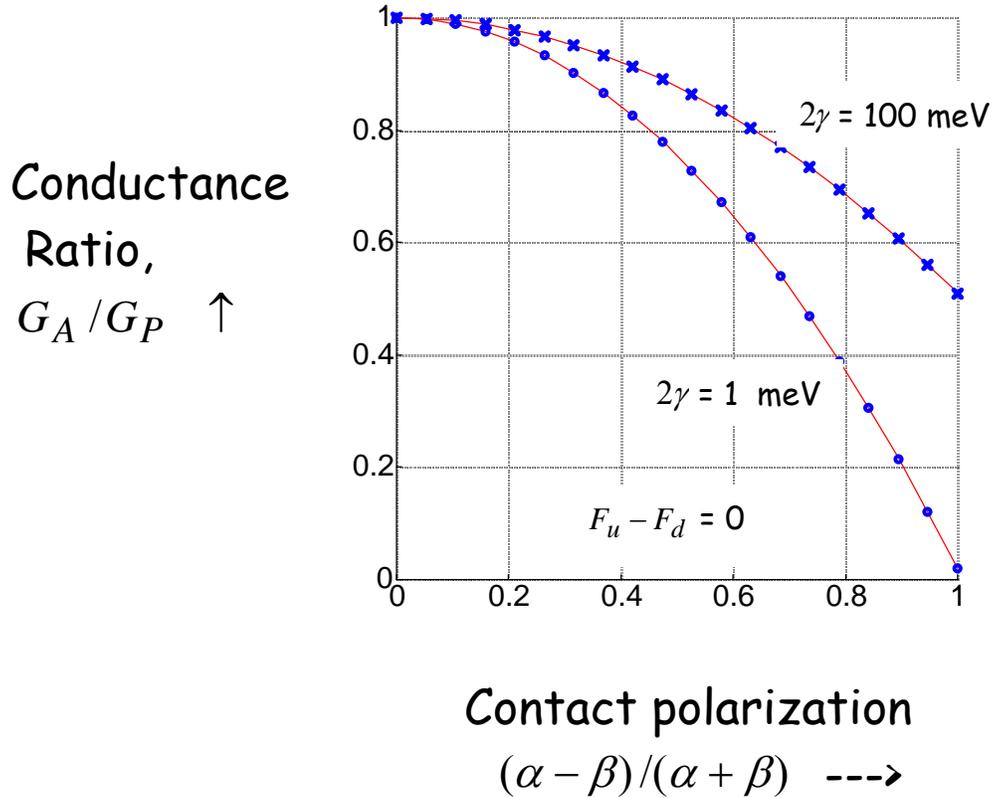

Fig.4.1. Conductance ratio, $G_A/G_P$ as a function of the contact polarization $P_c \equiv (\alpha-\beta)/(\alpha+\beta)$ calculated from the NEGF model (solid lines) compared with the result (Eq.(2.5)) from the heuristic equivalent circuit model with $2g_\gamma/(g_\alpha+g_\beta)$ = 1 ('x') and 1e-6 ('o'). Parameters used in NEGF model: $\varepsilon = 0, E_f$ = 0, $V_D$ = 100 meV, $\alpha+\beta$ = 100 meV and $2\gamma$ = 100 meV and 1 meV as indicated. Impurity spin is assumed unpolarized: $F_u - F_d$ = 0.

Not surprisingly, the numerical results obtained from the NEGF model match the prediction from the heuristic model in Section 2 quite well in this case (see Fig.4.1).

The results get more interesting when we consider polarized impurities with $F_d \neq F_u$, so that the quadratic term $\tilde{I}_\gamma$ is not zero. Fig.4.2 shows the current vs. voltage (I-V) for the same asymmetric spin valve device as in Fig.4.1, but with $F_u - F_d = 0, \pm 1$. With $F_u - F_d = 0$, the I-V curve passes through the origin as we would expect: Zero current for zero voltage. But with $F_u - F_d = \pm 1$, the curves are shifted up or down like that of an illuminated photodiode: The impurity spin acts like a current generator as shown in the equivalent circuit in Fig.2.2. The magnitude of this current generator is obtained by integrating $\tilde{I}_\gamma$ from Eq.(4.18c) over energy

$$I_\gamma = 2(e/h) \int dE \, J^2 N_I A^2 \, [f_d(1-f_u) + f_u(1-f_d)](F_u - F_d)$$

To estimate its magnitude we need $f_u$ and $f_d$, which is easy at equilibrium (zero applied voltage): $f_u = f_d = f_{eq}$. If we assume that the spectral functions are independent of energy in the energy range of interest, then we can write

$$\begin{aligned}[I_\gamma]_{V_D=0} &= 4(e/h)(F_u - F_d) J^2 N_I A^2 \int dE \, f_{eq}(1-f_{eq}) \\ &= 2(F_u - F_d) g_\gamma (k_B T / e) \end{aligned} \quad (4.20)$$

using Eq.(4.19b) and the identity

$$\int dE \, f_{eq}(1-f_{eq}) = k_B T \int dE \, (-\partial f_{eq}/\partial E) = k_B T$$

Using Eq.(4.20) in Eq.(2.6) we obtain an expression for the short-circuit current

$$I_{sc} = (F_u - F_d)(k_B T/e) \frac{2 g_\gamma (g_\alpha - g_\beta)}{2 g_\gamma + g_\alpha + g_\beta} \quad (4.21a)$$

Making use of the relation between the conductance and the broadening from Eqs.(4.19a,b) we can write

$$I_{sc} = 4(F_u - F_d) \frac{k_B T}{e} \frac{e^2}{h} \frac{2\gamma(\alpha - \beta)}{2\gamma + \alpha + \beta} A \qquad (4.21b)$$

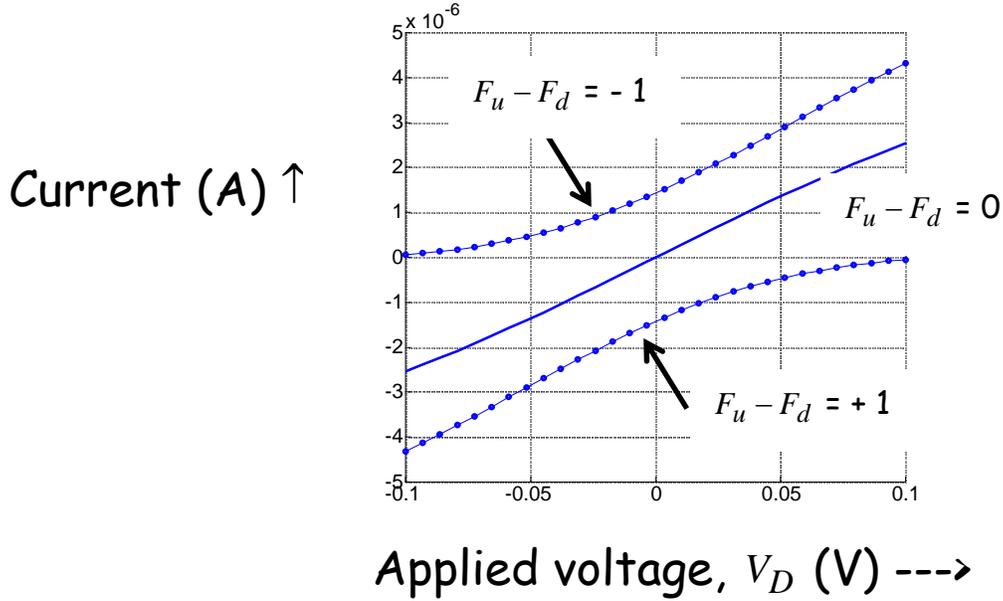

Fig.4.2. Current vs. voltage for the antiparallel spin-valve with contact polarization $P_c \equiv (\alpha - \beta)/(\alpha + \beta) = 1$ calculated from the NEGF model using the same parameters as in Fig.4.1: $\varepsilon = 0$, $E_f = 0$ with $\alpha + \beta = 100$ meV and $2\gamma = 100$ meV. Three values of impurity spin polarization are used as indicated: $F_u - F_d = 0, \pm 1$. Note that with $F_u - F_d = \pm 1$, current flows at zero bias: the energy comes at the expense of the entropy of the spins.

From Eq.(4.13) we could write approximately at energies around $E = \varepsilon$, $A \approx 4/(\alpha + \beta)$ so that

$$I_{sc} = 4(F_u - F_d) \frac{k_B T}{e} \frac{e^2}{h} \frac{2\gamma(\alpha - \beta)}{(2\gamma + \alpha + \beta)(\alpha + \beta)}$$

which agrees approximately with the numerical result in Fig.4.2. Note that a complete NEGF treatment requires the spectral function A to be calculated self-consistently. However, we have assumed it to have a fixed value of $A = \dfrac{\alpha+\beta}{(E-\varepsilon)^2 + ((\alpha+\beta)/2)^2}$ in obtaining the results shown in Fig.4.2, neglecting any change due to the additional broadening $\Gamma_u$ and $\Gamma_d$ (see Eq.(4.13)) from the spin-flip processes.

## 5. "Hot spin" effect

It is important to note that in the NEGF method described in the last section we assumed the impurity spin polarization to be fixed: The I-V characteristics change significantly as we go from unpolarized impurities with $F_u - F_d \equiv P_I = 0$ to polarized impurities with $P_I = \pm 1$ (see Fig.4.2). It is obvious that the impurity polarization $P_I$ could change through the interaction with the conduction electrons, but the implicit assumption is that the surroundings always restore the impurity to the equilibrium state, typically $P_I = 0$. Indeed this assumption is commonly made about all other "reservoirs" like phonons: Electrons may emit phonons as they traverse the channel, but the phonons quickly flow away and the phonon bath is always maintained at room temperature. People occasionally talk about "hot phonon" effects, but that usually requires large bias voltages. But my reason for choosing spins as an example is that they cannot flow away easily and so "hot spin" effects should be experimentally accessible under low bias conditions. Usually people work very hard to get rid of spin-flip impurities in spintronic devices and so hot spin effects are seldom encountered. Moreover, with the contacts in a parallel configuration, spin-flip interactions have minimal effect on the terminal current. By contrast, our "spin-charge transducer having antiparallel contacts is particularly suitable for observing the "hot spin" effect because the impurity polarization is bias-dependent and has a strong effect on teh terminal current.

***Dynamic polarization of impurities:*** The flow of current in the 'A' configuration creates a non-equilibrium distribution of spins in the channel electrons which in turn drives the impurity spin system into a non-equilibrium state. To model this dynamic polarization of impurities, we could use a set of rate equations of the form

$$(\hbar \, dF_u / dt) + \gamma_I (F_u - 0.5) = \gamma_d F_d - \gamma_u F_u \quad (5.1a)$$

$$(\hbar \, dF_d / dt) + \gamma_I (F_d - 0.5) = \gamma_u F_u - \gamma_d F_d \quad (5.1b)$$

where $\gamma_I$ represents the rate of impurity spin flipping through interactions other than the exchange interaction with the conduction electrons. The latter is described by the rate constants (see Eq.(4.18c))

$$\gamma_u = (2J^2/h) \int dE \, A_d A_u \, f_d(1 - f_u)$$

$$\text{and} \quad \gamma_d = (2J^2/h) \int dE \, A_u A_d \, f_u(1 - f_d) \quad (5.2)$$

Note that we have dropped the number of impurities $N_I$ appearing in the expression for the spin-flip current (Eq.(4.18c)) since we are writing a dynamical equation for one impurity. From Eqs.(5.1a,b) we can write for the impurity polarization $P_I \equiv F_u - F_d$,

$$(\hbar \, dP_I / dt) + (\gamma_I + \gamma_u + \gamma_d) P_I = \gamma_d - \gamma_u \quad (5.3)$$

The steady-state polarization is given by

$$P_I = (\gamma_d - \gamma_u)/(\gamma_I + \gamma_u + \gamma_d) \quad (5.4a)$$

If the rate $\gamma_I$ of relaxation of impurity spins by extraneous interactions is small enough ($\gamma_I \ll \gamma_u + \gamma_d$), then

$$P_I \cong (\gamma_d - \gamma_u)/(\gamma_u + \gamma_d) \quad (5.4b)$$

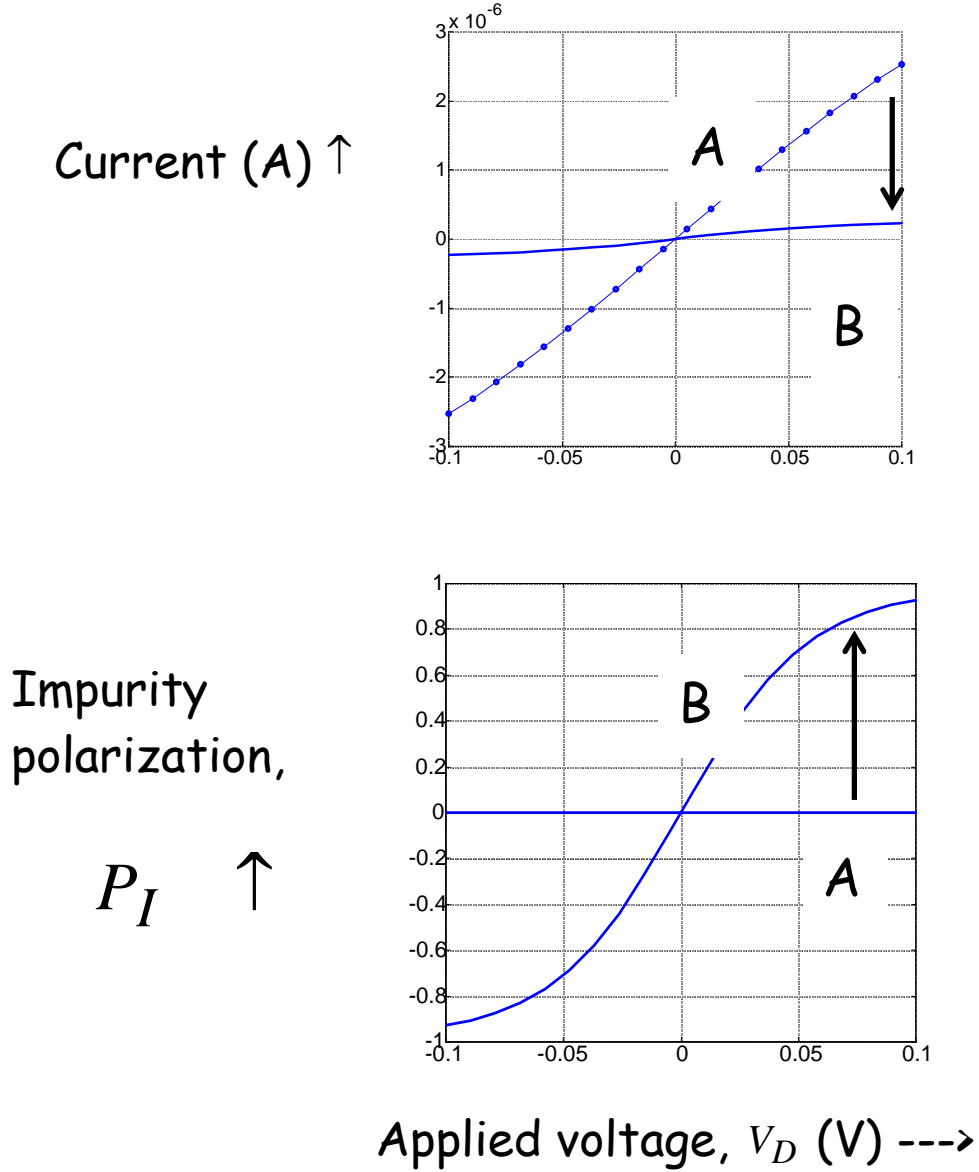

Fig.5.1. Current versus voltage for the same device as in Fig.4.2 but with the impurity spin polarization calculated self-consistently from Eq.(5.4a) assuming a relaxation rate $\gamma_I$ = 1 meV = 0.01 $\gamma$ = 0.01 ($\alpha+\beta$). The curve marked A shows the initial condition with $P_I = F_u - F_d = 0$, while the curve marked B shows the condition after $P_I$ has reached its steady-state value.

Fig.5.1a shows the current-voltage characteristics of an antiparallel spin valve with the same parameters as in Section 4, with a contact polarization $P_c \equiv (\alpha - \beta)/(\alpha + \beta)$ equal to one. The solid curve shows the result if we assume a fixed impurity polarization $P_I$ equal to zero (same as Fig.4.2). But if we solve for the impurity polarization ***self-consistently*** with Eq.(5.3), assuming an impurity relaxation rate of $\gamma_I = 1$ meV, we see a dramatic reduction in the current (Fig.5.1a, curve with 'o's) because the impurities get polarized (see Fig.5.1b) and stop acting as spin-flip scatterers. This effect would not be observed if we were using contacts in the parallel configuration. The effect would be less dramatic if the contact polarization $P_c \equiv (\alpha - \beta)/(\alpha + \beta)$ were less than one: we could write approximately for the steady-state polarization

$$P_I = \langle f_u(1-f_d)/f_d(1-f_u) \rangle$$
$$\approx (\alpha^2 - \beta^2)/(\alpha^2 + \beta^2) \tag{5.5}$$

The example we have shown corresponds to setting $\beta$ equal to zero, which requires the use of half-metallic contacts. But with less efficient contacts the effect should still be observable: One should see the I-V characteristic evolve from A to B (see Fig.5.1) as we approach steady-state.

***Time constant for impurity polarization:*** What is the time constant for this evolution? Or in other words, how long does it take to polarize the impurities and change the I-V curve from A to B? The rate is determined by $(\gamma_d + \gamma_u)$, (assuming it is much greater than $\gamma_I$) which can be estimated from Eq.(5.2) assuming zero voltage so that $f_u = f_d = f_{eq}$:

$$[\gamma_u + \gamma_d]_{V_D=0} = (4J^2 A^2/h) k_B T = 2\gamma A k_B T/N_I \tag{5.6}$$

making use of Eq.(4.19b). This shows that the rate of impurity polarization is equal to the spin-flip rate $\gamma$ multiplied by $(2A k_B T/N_I)$: this factor can be interpreted as the ratio of the number of conduction electrons $2A k_B T$ involved in spin-flip processes to the number of impurity spins $N_I$ with which they interact.

***Concluding remarks:*** To summarize, we have described a "spin-charge transducer" which is simply a ***spin valve with anti-parallel contacts*** having ***spin-flip impurities***. What makes this device useful conceptually (perhaps even practically) is the sensitivity of the terminal current to the degree of spin-flip scattering – something that is absent or at least greatly diluted in a similar device with parallel contacts. The terminal current thus acts as a sensitive probe for the degree of spin-flip scattering in this device. As long as external sources can maintain the impurities in an unpolarized state they act as spin-flip scatterers and a large current flows through the device (curve A of Fig.5.1). But if the rate of information erasure from the impurities cannot keep up with the information input from the electrons, the impurities will cease to act as dephasing agents and the current will be reduced (curve B of Fig.5.1).

Usually spin-flip impurities are considered deleterious to the operation of spintronic devices and one tries very hard to eliminate them. As such this device may sound purely academic. But one can conceive of memory devices where information is "written" in the form of impurity spin polarization through the hot spin effect and "read" through their effect on the terminal current. In this article, however, I have used this "spin-charge transducer" simply to illustrate a conceptual point, namely the importance of efficient "information erasure" in any dephasing process. Whether this gedanken device will find any use as a real experimental tool remains to be seen.

**Acknowledgements**

This work was supported by the Army Research Office (ARO) and the Defense Advanced Research Projects Agency (DARPA).

*Supriyo Datta, Purdue University*

   

**Appendix A: Dephasing in the NEGF formalism**

As I mentioned in the text dephasing processes are described within the NEGF formalism through the inscattering and outscattering functions which have to be calculated self-consistently from the correlation functions:

$$[\Sigma_s^{in}]_{ij} = \sum_{k,l} D_{ij;kl}^n G_{kl}^n \tag{A.1a}$$

$$[\Sigma_s^{out}]_{ij} = \sum_{k,l} D_{ij;kl}^p G_{kl}^p \tag{A.1b}$$

In this Appendix, I will describe how the functions $D^n$ and $D^p$ can be obtained, given the Hamiltonian

$$H = J\vec{\sigma}.\vec{S} \tag{A.2}$$

that is responsible for the dephasing interaction. The standard NEGF treatment usually involves second quantized operators and advanced concepts like the Keldysh contour. This treatment has been applied to electron-phonon and electron-electron scattering [8] problems, but we have not seen it applied to spin dephasing problems. We believe the correct results at least to first order (the so-called self-consistent Born approximation) can be obtained from a simple wavefunction based argument (Chapter 9, Ref.[1a]) described below.

A dephasing interaction like Eq.(A.2) involves both the electronic and the reservoir (in this case the impurity spin) coordinates and gives rise to a matrix element from an initial state (k,B) to a final state (i,A) where we are using lower case alphabets (like i, k) for electronic spin and upper case alphabets (like A, B) for the impurity spin. The source terms in (i,A) and (j,A) due to initial wavefunctions in (k,B) and (l,B) are written as

$$\tilde{S}_{iA} = H_{iA;kB} \psi_{kB} \quad \textbf{and} \quad \tilde{S}_{jA} = H_{jA;lB} \psi_{lB} \tag{A.3}$$

respectively. The inscattering function represents the correlation of the source terms:

$$\Sigma_{ij}^{in} = \sum_A \tilde{S}_{iA} \tilde{S}_{jA}^* = \sum_A H_{iA;kB} H_{jA;lB}^* \psi_{kB} \psi_{lB}^*$$

If $P_B$ is the probability that the impurity is in state 'B' (we assume that the states A, B are chosen such that the impurity density matrix is diagonal) then we can write ($G^n$: electron correlation function)

$$\psi_{kB} \psi_{lB}^* = P_B G_{kl}^n$$

so that making use of the fact that 'H' is Hermitian, we have

$$\Sigma_{ij}^{in} = \sum_{A,B} H_{iA;kB} H_{lB;jA} P_B G_{kl}^n$$

Comparing with Eq.(A.1a) we obtain

$$D_{ij;kl}^n = \sum_{A,B} H_{iA;kB} P_B H_{lB;jA} = Trace\big([H]_{ik} [\rho] [H]_{lj}\big)$$

This allows us to write $D^n$ in a general matrix form in the impurity subspace that can be used even if the impurity density matrix $\rho$ is not diagonal. We can obtain $D^p$ using similar arguments in terms of hole wavefunctions which can be viewed as the complex conjugate of the corresponding electron wavefunction:

$$D_{ij;kl}^n = N_I \, Trace\big([\rho][H]_{lj} [H]_{ik}\big) \qquad (A.4a)$$
$$D_{ij;kl}^p = N_I \, Trace\big([\rho][H]_{ik} [H]_{lj}\big) \qquad (A.4b)$$

$N_I$ being the total number of impurities.

***Evaluating $D^n$ and $D^p$ using Eqs. (A.4a,b):*** First we write the interaction Hamiltonian in Eq.(A.2) in the electron spin subspace:

$$H/J \;=\; \begin{pmatrix} & 1 & 2 \\ 1 & S_z & S_- \\ 2 & S_+ & -S_z \end{pmatrix} \tag{A.5}$$

where '$\vec{S}$' is a (2x2) Pauli spin matrix in the impurity spin subspace:

$$S_z \;=\; \begin{pmatrix} 1 & 0 \\ 0 & -1 \end{pmatrix} \tag{A.6a}$$

$$S_+ \equiv S_x + iS_y \;=\; \begin{pmatrix} 0 & 2 \\ 0 & 0 \end{pmatrix} \tag{A.6b}$$

$$S_- \equiv S_x - iS_y \;=\; \begin{pmatrix} 0 & 0 \\ 2 & 0 \end{pmatrix} \tag{A.6c}$$

Using Eq.(A.5) we obtain for the quantities appearing in Eqs.(A.4a,b):

$$H_{lj}\, H_{ik}/J^2 \;=\; \begin{array}{c} k,l \to \\ i,j \downarrow \end{array} \begin{pmatrix} & 11 & 22 & 12 & 21 \\ 11 & S_z^2 & S_+ S_- & S_+ S_z & S_z S_- \\ 22 & S_- S_+ & S_z^2 & -S_z S_+ & -S_- S_z \\ 12 & S_- S_z & -S_z S_- & -S_z^2 & S_-^2 \\ 21 & S_z S_+ & -S_+ S_z & S_+^2 & -S_z^2 \end{pmatrix} \tag{A.7a}$$

$$H_{ik}\, H_{lj}/J^2 \;=\; \begin{array}{c} k,l \to \\ i,j \downarrow \end{array} \begin{pmatrix} & 11 & 22 & 12 & 21 \\ 11 & S_z^2 & S_- S_+ & S_z S_+ & S_- S_z \\ 22 & S_+ S_- & S_z^2 & -S_+ S_z & -S_z S_- \\ 12 & S_z S_- & -S_- S_z & -S_z^2 & S_-^2 \\ 21 & S_+ S_z & -S_z S_+ & S_+^2 & -S_z^2 \end{pmatrix} \tag{A.7b}$$

Assuming an impurity density matrix of the form ($F_u + F_d = 1$)

$$\rho \equiv \begin{pmatrix} F_u & \Delta \\ \Delta^* & F_d \end{pmatrix} \quad \text{(A.8)}$$

we can evaluate the desired quantities $D^n$ and $D^p$ from Eqs.(A.5a,b):

$$D^n / J^2 N_I = \begin{array}{c} \phantom{11} \\ 11 \\ 22 \\ 12 \\ 21 \end{array} \begin{pmatrix} \begin{array}{cccc} 11 & 22 & 12 & 21 \end{array} \\ 1 & 4F_u & -2\Delta^* & -2\Delta \\ 4F_d & 1 & -2\Delta^* & -2\Delta \\ 2\Delta & 2\Delta & -1 & 0 \\ 2\Delta^* & 2\Delta^* & 0 & -1 \end{pmatrix} \quad \text{(A.9a)}$$

$$D^p / J^2 N_I = \begin{array}{c} \phantom{11} \\ 11 \\ 22 \\ 12 \\ 21 \end{array} \begin{pmatrix} \begin{array}{cccc} 11 & 22 & 12 & 21 \end{array} \\ 1 & 4F_d & 2\Delta^* & 2\Delta \\ 4F_u & 1 & 2\Delta^* & 2\Delta \\ -2\Delta & -2\Delta & -1 & 0 \\ -2\Delta^* & -2\Delta^* & 0 & -1 \end{pmatrix} \quad \text{(A.9b)}$$

Note that the top (2x2) segment of these matrices is identical to Eqs.(4.6a,b) used in the main paper. As we noted there, this segment can be understood from simple common sense arguments since the (11) and (22) elements are positive numbers that represent the occupation of levels 1 and 2. By contrast the (12) and (21) elements are in general complex numbers representing the phase correlation between levels 1 and 2 and it is not as easy to write down the corresponding elements of $D^n$ and $D^p$ from simple reasoning. However, these elements have not been needed in this paper since we have only discussed examples that do not involve off-diagonal terms of the correlation function.